\documentclass[12pt,a4paper]{article}
\usepackage{graphics}
\usepackage{amsfonts}

\newcommand{\preprint}[1]{\begin{table}[t]  
           \begin{flushright}               
           \begin{large}{#1}\end{large}     
           \end{flushright}                 
           \end{table}}                     

\newcommand{\LL}{{\cal L}}
\newcommand{\HH}{{\cal H}}
\newcommand{\be}{\begin{equation}}
\newcommand{\ee}{\end  {equation}}
\newcommand{\pa}{\partial}

\newtheorem{theorem}{Theorem}
\newtheorem{lemma}{Lemma}
\newtheorem{definition}{Definition}

\input{psfig}

\begin{document}

\begin{titlepage}

\preprint{hep-th/9811192\\TAUP-2536-98}

\vspace{2cm}

\begin{center}
{\bf\LARGE $Q \bar{Q}$ Potential from Strings in Curved Spacetime --  
          Classical Results
}\footnote{
Work supported in part by the US-Israel Binational Science
 Foundation,
by GIF - the German-Israeli Foundation for Scientific Research,
and by the Israel Science Foundation.
}\\
\vspace{1.5cm}
{\bf Y. Kinar  $\;\;$ E. Schreiber $\;\;$ J. Sonnenschein}\\
\vspace{.7cm}
{\em Raymond and Beverly Sackler Faculty of Exact Sciences\\
 School of Physics and Astronomy\\
 Tel Aviv University, Ramat Aviv, 69978, Israel\\
e-mail: \{yaronki,schreib,cobi\}\verb+@+post.tau.ac.il}
\end{center}
\vskip 0.61 cm

\begin{abstract}
We compute the 
leading behaviour of the quark anti--quark potential from 
a generalized Nambu--Goto action associated with a curved space--time
having an "extra dimension".
The extra dimension can be the 
radial coordinate in the AdS/CFT correspondence, 
the Liouville field in
Polyakov's approach, or an internal dimension in MQCD.
In particular, we derive the  condition for confinement,
and in the case it occurs we find the string tension and the 
correction to the linear potential.


\end{abstract}

\end{titlepage}

\baselineskip 18pt

\section { Introduction}
Recently, the idea of presenting the Wilson loop of non-abelian gauge
theories in terms of $<e^{-S}>$, where $S$ is a string worldsheet area, 
has undergone a Renaissance period. 
This type of construction  of the quark anti--quark 
potential has  emerged mainly in the framework of the 
novel gravity/gauge field duality \cite{Mal2} but also  in the context of
the M-theory description of QCD (MQCD) \cite{Wit2,HSZ,KSS1}  
and  in Polyakov's Liouville approach \cite{Polyakov}.

In the gravity/gauge duality approach, physical quantities of the 
boundary gauge theory are computed in terms of the bulk gravitational
properties. In particular, the string between the quark anti--quark pair
is not confined  to the four dimensional boundary but rather stretches
inside the five dimensional $AdS_5$ part of the  ten dimensional
space-time. A non-flat five dimensional space is also the picture
that Polyakov draws for the non-critical string  proposed as
a solution of the loop equation originating from  non--abelian 
gauge dynamics. In MQCD the QCD string translates into a  membrane M2  
ending on the five-brane
M5 which describes the super YM (or super QCD) degrees of freedom.
One of the M2 coordinates  is along a trajectory embedded in the 
coordinates transverse to the four dimensional  space time.  
Thus, a common  concept invoked in these "modern" calculations is the fact
that  the Wilson loop is a boundary  of the 
string world sheet which is embedded in  a higher (than four) dimensional 
space-time. 

The space-time metric  that associates with   these three setups
 is a diagonal one and is a function of 
only  one coordinate. It is the fifth 
coordinate of the $AdS_5$
(or its analogs in the non-extremal cases), 
the fifth direction of the M5 brane in MQCD, 
and the Liouville coordinate in Polyakov's approach.
Denoting this coordinate by $s$, the metric takes the 
following generic form
$$ ds^2 = G_{MN}dx^Mdx^N= -G_{00}(s) dt^2 + G_{x_{||}x_{||}}(s) dx^2_{||}
+ G_{ss}(s) ds^2
 + G_{x_{T}x_{T}}(s) dx^2_{T}$$
where $x_{||}$ are the $R^3$ ordinary space  coordinates
(or possible generalization to 
p-brane space coordinates in the gravity/gauge duality approach) and
$x_{T}$ are the coordinates transverse to the five dimensional space. 
This  case of three dimensional space can be easily generalized  
to $p$ dimensional
space with $x_T$ being the coordinates of the $8-p$ dimensional 
transverse space. 
Upon choosing the world sheet coordinates $\sigma=x$ and $\tau=t$ 
 and assuming translation invariance along $t$, 
 the Nambu--Goto string  action takes the form 
\begin{eqnarray}
S&=&\int d\sigma d\tau \sqrt{det[\pa_\alpha  X^M \pa_\beta X^N
G_{MN}]} \\
&=&T \cdot \int dx\sqrt{ G_{00}(s(x))G_{x_{||}x_{||}}(s(x)) + 
 G_{00}(s(x))G_{ss}(s(x))(\pa_x s)^2}
\end{eqnarray}
With $T$ the total time of the Wilson loop rectangle.
In spatial  Wilson loop calculations 
(see for example \cite{BISY2}) the two sides of
the Wilson rectangle are taken to be along space directions $(x,x')$.
In these cases it is convenient to choose $\tau=x'_{||}$ and therefore
$G_{00}$ has to be replaced with $G_{x_{||}'x_{||}'}$, and $T$ by $L'$,
the total length along $x'_{||}$ of the Wilson loop rectangle. 
For convenience we define now 
\begin{eqnarray}
f^2(s(x)) & \equiv & G_{00}(s(x))G_{x_{||}x_{||}}(s(x)) \\
g^2(s(x)) & \equiv & G_{00}(s(x))G_{ss}(s(x))
\end{eqnarray}
so that the Nambu--Goto action reads
\begin{equation}
\label{NGaction}
S=T \cdot \int dx\sqrt{ f^2(s(x)) + g^2(s(x)) (\pa_x s)^2}
\end{equation}
Naturally, this metric should be positive, and we can assume that the  
functions $f(s)$ and $g(s)$ are real and non negative.

To get familiarized with this form of the action we  write now the form that 
 $f^2(s(x))$ and $g^2(s(x))$  take  in  certain examples. 
\begin{itemize}
\item In the original  AdS/CFT case \cite{Mal2}, 
it is customary to use $U$ instead of $s$. The functions are 
\begin{eqnarray*}
f^2(U(x)) & = & (2\pi)^{-2} (U/R_{AdS})^4 \\
g^2(U(x)) & = & (2\pi)^{-2}
\end{eqnarray*}
where $R_{AdS}^4=4\pi g N$. 

\item In the supergravity setup corresponding to the "pure YM case" 
\cite{BISY2} one finds 
\begin{eqnarray*}
f^2(U(x)) & = & (2\pi)^{-2} (U/R_{AdS})^4 \\
g^2(U(x)) & = & (2\pi)^{-2} (1-(U_T/U)^4)^{-1}
\end{eqnarray*}
with $U_T$ related to the energy density.

\item In the near extremal AdS solution corresponding to field
theory  at finite temperature \cite{BISY1},
\begin{eqnarray*}
f^2(U(x)) & = & (2\pi)^{-2} (U/R_{AdS})^4 (1-(U_T/U)^4) \\
g^2(U(x)) & = & (2\pi)^{-2}
\end{eqnarray*}
when $U_T/(\pi R_{AdS}^2)$ is the Hawking temperature.

\item In the "pure YM" theory corresponding to the supergravity
  solution of rotating branes \cite{CORT}, the functions are 
(in M theory units)
\begin{eqnarray*}
f^2(U(x)) & = & C \frac{U^6}{U_0^4} \Delta \\
g^2(U(x)) & = & \frac{C U^2 \Delta}{1-a^4/U^4 - U_0^6/U^6}
\end{eqnarray*}
where $\Delta = 1 - a^4 \cos^2 \theta / U^4$,   
$\;\, a$ parameterizes the angular momentum of the brane 
(whose rotation is limited to a single plane), $U_0$ is the location 
of the horizon, $\theta$ is a coordinate of the internal space 
(which is asymptotically $S^4$), and $C$ is a constant with the 
correct dimensions. 

\item For the $SU(2)$  case of MQCD, it was realized \cite{KSS1} that 
\begin{eqnarray*}
f^2(s(x)) & = & 8 \zeta \cosh(s/R) \\
g^2(s(x)) & = & 8 \zeta \cosh(s/R)  
\end{eqnarray*}
with $R \sim \Lambda_{QCD}^{-1}$ (the radius of the 11-th dimension)
and $\zeta \sim \Lambda_{QCD}^4$ parameterizing the M-theory curve.

\item In Polyakov's approach, the fifth coordinate is the Liouville
  field $\phi$. In that case \cite{Polyakov}
\begin{eqnarray*}
f^2(\phi(x)) & = & a^4(\phi) \\
g^2(\phi(x)) & = & a^2(\phi)
\end{eqnarray*}
where $a(\phi)$ is determined by   conformal invariance.  
\end{itemize}

In this note we derive the  quark anti--quark potential, namely the Wilson
loop, that associates with the Nambu--Goto action (\ref{NGaction}).
Our analysis is based on the classical equations of motion and does not 
include quantum fluctuations \cite{Luscher}. 
Our main result, which is stated below in a rigorous way, 
is that (assuming without loss of generality that $f(s)$ has a minimum or 
$g(s)$ diverges at $s=0$)
{ \it confinement occurs if and only if $f(0)>0$ and the 
corresponding string tension is  $f(0)$}. 
In addition we show that  when $f(0) = 0$, the potential behaves 
asymptotically as a (negative) power of the separation of the quark
and anti--quark, and we find the exact power and coefficient. 
When $f(0) \ne 0$, apart from the linear potential and a constant term,
we find the form of the next correction. For the critical case when the 
minimum of $f(s)$ is just deep enough (or the divergence of $g(s)$ is just
strong enough) to allow the separation to diverge as the string approaches
the minimum, the correction is exponentially small. 
At the non critical cases, the correction is power--like. In both cases,
we explicitly find the relevant constants.

In section 2  we present the classical analysis of the action, 
we compute the quark anti--quark potential  and state
 the main result of the paper. 
In Section 3  we present a rigorous   proof of our statement.
Section 4 is devoted to  several examples to which we apply our
general result. In section 5 we deal with a variant of our main analysis.
In section 6 we give summary and conclusions. 

\section{The quark anti--quark  potential}

We shall write the Lagrangian density (relative to $x$) corresponding to
the general Nambu--Goto action (\ref{NGaction}). We take the Lagrangian without
the factor $T$, and therefore the action derived from it represents the 
quark anti--quark potential.
\be
\label{Lgeneral}
\LL(s,s') = \sqrt{f^2(s) + g^2(s) s'^2}
\ee
As $S = T \cdot \int \LL \, dx$ is dimensionless, we see that in our
formulation, $\LL$ has dimensions of $\mbox{mass}^2$. 

The conjugate momentum is 
\be
\label{momentum}
p = \frac{\delta \LL}{\delta s'} = \frac{g^2(s) s'}{\sqrt{f^2(s) + g^2(s) s'^2}}
\ee

and therefore
\be
\label{Hs}
\HH(s,p) = p \cdot s' - \LL(s,s'(s,p)) =
\frac{-f^2(s)}{\sqrt{f^2(s) + g^2(s) s'^2}} = -\frac{f^2(s)}{\LL}
\ee

As the Hamiltonian $\HH$ does not depend explicitly on $x$, its value
is a constant of motion. We shall deal with the case in which $s(x)$
is an even function, and therefore there is a minimal value 
$s_0 = s(0)$ for which $s'(0) = 0$. At that point, we see
(\ref{momentum}) that $p = 0$ also. The constant of motion is,
therefore, 
\be
\label{Hs0}
\HH(s_0,0) = -f(s_0)
\ee

From (\ref{Hs},\ref{Hs0}) we can express the Lagrangian without
taking recourse of $g(s)$:
\be
\LL = \frac{f^2(s)}{f(s_0)}
\ee 
and we can also extract the differential equation of the geodesic
line:
\begin{equation}
\label{sofx}
\frac{d s}{d x} = \pm \frac{f(s)}{g(s)} \cdot
\frac{\sqrt{f^2(s)-f^2(s_0)}}{f(s_0)}
\end{equation}

The distance (in the ordinary space) between two "quarks" situated
at $s = s_1$ is, therefore, 
\be
\label{lgeneral}
l = \int dx = \int \left(\frac{d s}{d x}\right)^{-1} ds = 
2 \int_{s_0}^{s_1} \frac{g(s)}{f(s)} \frac{f(s_0)}{\sqrt{f^2(s)-f^2(s_0)}} ds 
\ee

The energy of the configuration is the length of the string according
to the metric (\ref{Lgeneral})
\begin{eqnarray}
\label{Eprime}
E' & = & \int \LL dx = \int \left(\frac{d s}{d x}\right)^{-1} \LL ds =
2 \int_{s_0}^{s_1} \frac{g(s)}{f(s)}
\frac{f^2(s)}{\sqrt{f^2(s)-f^2(s_0)}} ds \\
   & = & f(s_0) \cdot l + 2 \int_{s_0}^{s_1} \frac{g(s)}{f(s)}\sqrt{f^2(s)-f^2(s_0)} ds 
\end{eqnarray}

In order to get the potential between the "quarks", we have to
subtract the masses of the two quarks. The masses are independent of $s_0$,
and their subtraction is needed only to regulate the singularities in 
(\ref{Eprime}). Other subtraction schemes (e.g the one used in \cite{BISY2})
are possible, and would result in a constant shift of the potential. As we 
analyze the behaviour of the potential only for large  separations, 
the constant term 
has no direct physical meaning. 
Nevertheless, in our setting it is most natural to take as the "bare" quark
a straight string with a constant value of $x$, stretching from $s = 0$ to
$s = s_1$. Each quark has, then,  a mass of
\be
m_q = \int_0^{s_1} g(s) ds
\ee
Moreover, we shall find that in this
subtraction scheme, the sign of the potential is naturally related to the
globality of the minimum.
We would like further to comment that this choice sets a limit on the 
permitted divergence of $g(s)$ as $s \rightarrow 0$.   

For the potential we get, therefore,
\begin{eqnarray}
\label{Egeneral}
E & = & f(s_0) \cdot l + 2 \int_{s_0}^{s_1} \frac{g(s)}{f(s)} \;
\left(\sqrt{f^2(s)-f^2(s_0)} - f(s)\right) ds - 2 \int_0^{s_0} g(s) ds \\
\label{Egeneral2}
  & \equiv & f(s_0) \cdot l - 2 K(s_0)
\end{eqnarray}

with
\footnote{$K$ is similar, but not identical,
to $-J$ in \cite{KSS1}.} 
\be
\label{Ks0}
K(s_0) = \int_{s_0}^{s_1} \frac{g(s)}{f(s)} \; 
\left(f(s) - \sqrt{f^2(s)-f^2(s_0)}\right) ds + \int_0^{s_0} g(s) ds
\ee
We define also $\kappa = K(0)$.

The geometric picture we are depicting is quite simple. Under suitable 
assumptions, The geodesic line can not pass a value of $s$ for which 
either $f(s)$ has a minimum or $g(s)$ diverges. without loss of generality, 
we can take this value of $s$ to be $0$. For large $l$, then, the geodesic 
line lies, for most of its part, very close to $s = 0$. The first term of the
expression (\ref{Egeneral2}) tends, then, to $f(0) \cdot l$, while the second
and third ones are $O(1)$ (bounded by a constant in $l$). We therefore get
that there is confinement (i.e linear potential) for the metric 
(\ref{Lgeneral}), if and only if $f(0) \ne 0$, and then $f(0)$ is the 
string tension. Note that $f(s)$ has the same dimensionality as $\LL$, 
i.e $\mbox{mass}^2$, which is indeed the dimensionality of string tension. 
In the following section we prove all of our claims.

\section {The proof}
Our proof of the statement made above on the functional dependence
of $E(l)$ includes the followings steps. We first
prove in theorem \ref{canoe} that $s_0$ is a monotonic decreasing function
of $l$. In theorems \ref{technical1} and \ref{technical2} we write down 
expressions for the asymptotic behavior of $l(s_0)$ and $K(s_0)$.  Finally the 
asymptotics of $E(l)$ are derived in theorem \ref{main}. 

\subsection{Monotonicity of $l(s_0)$}

The solution (\ref{sofx}) of the Euler--Lagrange equations does not have 
to be the global minimum. It is possible that the global minimum is non 
differentiable, and therefore (\ref{sofx}) is not sensitive to it. However,
from the triangle inequality it is clear that a "corner" can occur in a
minimum--action function only if there is a direction in which the metric
is zero. Therefore, it is possible in our setting only if $f(0) = 0$ and
the string reaches $s=0$. In that case it is clear that the best configuration
is two bare quarks connected with a string segment on $s = 0$ (which does 
not "cost" any energy). This configuration, in our conventions, has zero
energy (as we subtract exactly the masses of the two quarks). Therefore we 
conclude that if the energy of the string describing the Euler--Lagrange 
geodesic has $E < 0$, it is the true solution (global minimum). 

\begin{figure}[h!]
\begin{center}
\resizebox{0.8\textwidth}{!}{\includegraphics{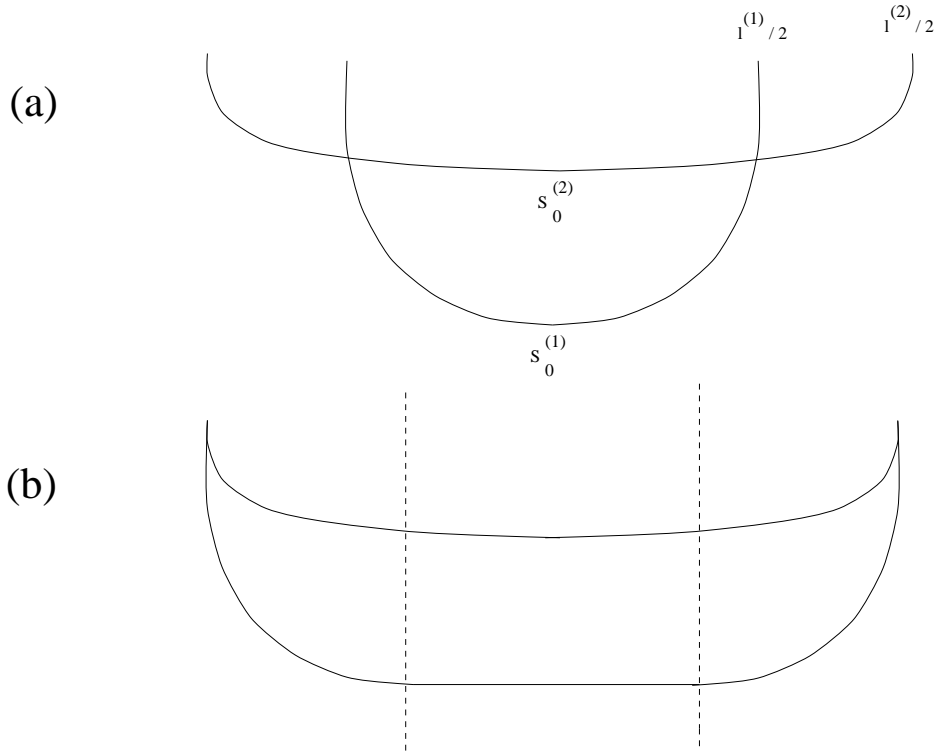}}
\end{center}
\caption{(a) the geodesic lines $G^{(1)},G^{(2)}$. 
         (b) $G^{(2)}$ and the new line having smaller energy.}
\label{fig:proof}
\end{figure}

\begin{theorem}
\label{canoe}
Let $f(s),g(s) > 0$ for $s > 0$, and let $f(s)$ be monotone increasing.
Assume that $s(x)$ described by (\ref{sofx}) has the global minimum value
of $E$. Then $s_0$ is monotone decreasing as a function of $l$. 
\end{theorem}
{\rm Proof:} Assume the contrary. Then there are two intersecting 
geodesic lines $G^{(1)},G^{(2)}$ 
with $s_0^{(1)} < s_0^{(2)} , l^{(1)} < l^{(2)}$ (see figure 
\ref{fig:proof}a). 

We shall now build a new line for $l^{(2)}$.
The new line will consist of the two halves of $G^{(1)}$, separated so 
they span the distance $l^{(2)}$, with a straight segment in the middle,
(lying at $s = s_0^{(1)}$ and of length $l^{(2)} - l^{(1)}$). Obviously,
$s' = 0$ for that straight segment. From our assumption 
$s_0^{(1)} < s_0^{(2)}$, this segment has a smaller value of $s$ than any
point of the corresponding curved segment of $G^{(2)}$. As $f(s)$ is 
monotone increasing, $f(s)$ is smaller for the straight segment than
for any point on the curved one. Therefore, $\LL$ is
also smaller (\ref{Lgeneral}) for the straight segment.

The total energy of the curved parts of the new line is smaller than that
of the corresponding parts of $G^{(2)}$, for otherwise those latter parts 
combined would give a line for $l^{(1)}$ with energy smaller than that of  
$G^{(1)}$ (see figure \ref{fig:proof}b). 
(Although that combined line is not differentiable at 
$s = s_0^{(2)}$, it can be smoothed so that its energy is changed by no more 
than $\epsilon$ for every $\epsilon > 0$.)
Therefore, the new line has a smaller energy then the assumed 
geodesic global minimum for $G^{(2)}$, which is a contradiction.

\subsection{Asymptotics of $l(s_0)$ and $K(s_0)$}

Let us assume that $f(s)$ has a minimum at $s = 0$.
we claim that under suitable assumptions, the first term of the
expression (\ref{Egeneral}) tends to $f(0) \cdot l$, while the second
and third ones are $O(1)$ (bounded by a constant in $l$). We therefore get
that there is confinement (i.e linear potential) for the metric 
(\ref{Lgeneral}), if and only if $f(0) \ne 0$ (and then $f(0)$ is the 
string tension). We shall now make those arguments more exact.

First, we need to state some preliminaries. 
In what follows, we denote by the symbol~$\sim$ 
that two functions behave alike, up to a non zero 
multiplicative constant. That is,
\begin{definition}
$h_1(s) \sim h_2(s)$ in a region if there exist constants
$a,A > 0$ such that for all $s$ in that region,
$a h_2(s) \le h_1(s) \le A h_2(s) $ 
\end{definition}
Obviously, the relation $\sim$ is an equivalence relation.

\begin{lemma}
\label{Cbehaviour}
Define 
\begin{eqnarray}
C_{n,m}(\tilde{y})  & = & \int_1^{\tilde{y}} \frac{dy}{y^n \sqrt{1 - y^{-m}}}\\
C'_{n,m}(\tilde{y}) & = & \int_1^{\tilde{y}} 
                                \frac{2(1 - \sqrt{1 - y^{-m}})\,dy}{y^{n-m}}
\end{eqnarray}

Then, if $m > 0$,
\begin{enumerate}
\item \label{Cgt1} if $n > 1$, then $0 < C_{n,m}(\infty) < \infty$, and
$C_{n,m}(\tilde{y}) = C_{n,m}(\infty) - O(\tilde{y}^{-(n-1)})$. 
\item \label{Ceq1} if $n = 1$, then 
$C_{n,m}(\tilde{y}) = \log \tilde{y} + O(1)$.

\item \label{Clt1} if $0 < n < 1$, then 
$C_{n,m}(\tilde{y}) = \frac{1}{1-n} \tilde{y}^{1-n} + O(1) 
                                                    + O(\tilde{y}^{1-n-m})$.
\end{enumerate}
Similar relations hold for $C'_{n,m}(\tilde{y})$.
\end{lemma}
{\rm Proof:} the lower limit of the integrals does not diverge. At the upper
limit, the integrands behave as $y^{-n}$, and the principal behaviours follow.
The corrections for assertion \ref{Cgt1} follow from the boundaries 
of the integral of $y^{-n}$, and from the expansion of $1/\sqrt{1 - y^{-m}}$. 
As for assertions \ref{Ceq1} and \ref{Clt1}, they follow from separating
$y^{-n}$, whose integral gives the divergence, from the integrands.

\begin{definition}
\be
D_{n,m} \equiv \frac{1}{m-n+1} + \frac{1}{2} C'_{n,m}(\infty) - C_{n,m}(\infty) 
\ee
\end{definition}

\begin{lemma}
\label{Dpositive}
If $k>0$ and $-1<j<k-1$ then $D_{2k-j,2k}>0$
\end{lemma}
{\rm Proof:} First we show that for a given value of $m$, $D_{n,m}$
is a monotone increasing function of $n$ (when both $m$ and $n$ are in
the specified range). 
In this range, $D_{n,m}$ may be written as  
\be
D_{n,m} = \int_1^\infty \left( y^{n-m-2} + y^{m-n} \:\frac{\sqrt{1-y^{-m}}-1}
                                                 {\sqrt{1-y^{-m}}} \right) dy  
\ee
Differentiating with respect to $n$ gives 
\be 
\frac{\pa D_{n,m}}{\pa n} = \int_1^\infty dy 
                            \frac{y^{m-n} \log y}{\sqrt{1-y^{-m}}} \left( \sqrt{1-y^{-m}} \:\,(y^{2(n-m-1)} - 1)+1 \right) 
\ee
The last integrand is obviously positive for $1<y<\infty$ and we find
that the derivative is positive. 

To complete the proof, we look at $D_{n,m}$ for the minimal value of
$n$, that is, the maximal value of $j$, which is $k-1$.
\begin{eqnarray}
D_{k+1,2k} & = & \int_1^\infty 
                 \left( y^{-(k+1)} + y^{k-1} \: \frac{\sqrt{1-y^{-2k}}- 1}
                                {\sqrt{1-y^{-2k}}} \right) dy \nonumber \\
           & = & \left. - (\sqrt{1-y^{-2k}} ( y^k - \sqrt{y^{2k}-1}))/k 
                 \right|_1^\infty \\
           & = & 0 \nonumber
\end{eqnarray}

\vspace{12pt}

We shall investigate the behaviour of $l$ as $s_0$ approaches 
(without loss of generality) the value $0$, assuming it is finite. 
We shall find that this behaviour is governed by the expansions of the 
functions $f(s)$ and $g(s)$. As the functions used in (\ref{Lgeneral}) are
$f^2(s)$ and $g^2(s)$, it may very well happen that the powers of the leading
terms in those expansions will be half integers. Therefore, we do not assume
that those powers are integer. 
Those considerations serve as motivations for the conditions in the 
following theorem.

\begin{theorem}
\label{technical1}
Let $f(s)$ be a function for $s > 0$, such that for $s$ 
close enough to $0$,
\be
\label{expansionf}
f(s) = f(0) + a_k s^k + O(s^{k+1})
\ee 
with $k > 0 \;,\; a_k > 0$. 
Let $g(s)$ be such that for $s$ close enough to $0$,
\be
\label{expansiong}
g(s) = b_j s^j + O(s^{j+1})
\ee
with $b_j > 0$. 
Assume also that $f(s),g(s) \ge 0$ for $0 < s < \infty$.

Take any $0 < s_1 \le \infty$ such that $l = l(s_0)$, as in (\ref{lgeneral}),
converges for $0 < s_0 < s_1$. Then, as $s_0$ tends to $0$ from above,
\begin{enumerate}
\item if $f(0) \ne 0$,
\begin{enumerate}
\item \label{l:ne0lt} if $k < 2(j+1)$, then $l$ is bounded.
\item \label{l:ne0eq} if $k = 2(j+1)$, then 
$l = -\frac{2 b_j}{\sqrt{2 f(0) a_k}} \log s_0 + \lambda$, with
   \begin{enumerate}
     \item \label{l:ne0eq:loglog}    $\lambda  =  O(\log (-\log s_0))$
     \item \label{l:ne0eq:logloglog} $\lambda \le O(\log \log (-\log s_0))$.
   \end{enumerate}
\item \label{l:ne0gt} if $k > 2(j+1)$, then 
$l = \frac{2 b_j}{\sqrt{2 f(0) a_k}} C_{k/2-j,k}(\infty) s_0^{-(k/2-j-1)} 
                                + O(s_0^{-(k/2-j-1) + \frac{k/2-j-1}{k/2-j}})$.
\end{enumerate}
In particular, $l$ diverges for $s_0 \rightarrow 0$ if and only if
$k \ge 2(j+1)$. 
\item if $f(0) = 0$,
\begin{enumerate}
\item \label{l:eq0lt} if $k < (j+1)/2$, then $l = O(s_0^k)$.
\item \label{l:eq0eq} if $k = (j+1)/2$, then $l \sim -s_0^k \log s_0$.
\item \label{l:eq0gt} if $k > (j+1)/2$, then 
$l = \frac{2 b_j}{a_k} C_{2k-j,2k}(\infty) s_0^{j+1-k}  
                     + O(s_0^{j+1-k + \frac{2k-j-1}{2k-j}})$.
\end{enumerate}
In particular, if $j>-1$ then $l$ diverges for $s_0 \rightarrow 0$ if and only if $k > j+1$.
\end{enumerate}
\end{theorem}
{\rm Proof:} We choose some $\tilde{s}$ such that the expansions 
(\ref{expansionf},\ref{expansiong}) are valid, and $f(s)$ is increasing,
 for $0 \le s \le \tilde{s}$.
We shall separate the integration range for $l$ into two parts - 
below $\tilde{s}$ and above it.

The integral in the second range is 
\be
\Delta l \equiv 2 \int_{\tilde{s}}^{s_1} \frac{g(s)}{f(s)} \frac{f(s_0)}{\sqrt{f^2(s)-f^2(s_0)}} ds \le \frac{f(s_0)}{f(\tilde{s})} \cdot 
2 \int_{\tilde{s}}^{s_1} \frac{g(s)}{f(s)} \frac{f(\tilde{s})}{\sqrt{f^2(s)-f^2(\tilde{s})}} ds = \frac{f(s_0)}{f(\tilde{s})} \cdot l(\tilde{s})
\ee
so $\Delta l = O(l(\tilde{s}))$ and $\Delta l \le l(\tilde{s})$ 
for $f(0) \ne 0$, and 
$\Delta l = O((s_0/\tilde{s})^k l(\tilde{s}))$ for $f(0) = 0$. 

Now we look at the integral in the first range, which can be the cause
of the divergence of $l(s_0)$.
Let us assume first that $f(0) \ne 0$. Then 
$f^2(s) - f^2(s_0) = (1 + O(\tilde{s})) 2 f(0) a_k (s^k - s_0^k)$ there, so
\begin{eqnarray}
\nonumber
2 \int_{s_0}^{\tilde{s}} \frac{g(s)}{f(s)} \frac{f(s_0)}{\sqrt{f^2(s)-f^2(s_0)}} ds & = & (1 + O(\tilde{s})) \frac{2 b_j}{\sqrt{2 f(0) a_k}} \cdot \int_{s_0}^{\tilde{s}} \frac{s^j ds}{\sqrt{s^k - s_0^k}} \\
\nonumber
& = & (1 + O(\tilde{s})) \frac{2 b_j}{\sqrt{2 f(0) a_k}} s_0^{j+1-(k/2)} \cdot
\int_1^{\tilde{s}/s_0} \frac{dy}{y^{k/2-j} \sqrt{1 - y^{-k}}} \\
& = & (1 + O(\tilde{s})) \frac{2 b_j}{\sqrt{2 f(0) a_k}} s_0^{j+1-(k/2)} \cdot
C_{k/2-j,k}(\tilde{s}/s_0)
\end{eqnarray}
with $y = s/s_0$. Substituting the behaviour of $C_{k/2-j,k}$ and
taking $\tilde{s}$ fixed, we get that $l$ is bounded for $k < 2(j+1)$,
$l \sim -\log s_0$ for $k = 2(j+1)$, and $l \sim s_0^{j+1-k/2}$ for 
$k > 2(j+1)$. 

To get sharper results, we now let $\tilde{s}$ vary with
$s_0$. By choosing $\tilde{s} = s_0^{(k/2-j-1)/(k/2-j)}$ 
when $k > 2(j+1)$ we prove assertion \ref{l:ne0gt}.
By choosing $\tilde{s} = -1/\log s_0$ when $k = 2(j+1)$ and using 
$\Delta l = O(l(\tilde{s}))$ we prove assertion \ref{l:ne0eq:loglog}.
Using that assertion for $l(\tilde{s})$ and $\Delta l \le l(\tilde{s})$,
again with $\tilde{s} = -1/\log s_0$, we prove assertion 
\ref{l:ne0eq:logloglog}.

Let us now assume that $f(0) = 0$. Now, 
$f^2(s) - f^2(s_0) = (1 + O(\tilde{s})) a_k^2 (s^{2k} - s_0^{2k})$ for 
$s_0 \le s \le \tilde{s}$, and 
\begin{eqnarray}
2 \int_{s_0}^{\tilde{s}} \frac{g(s)}{f(s)} \frac{f(s_0)}{\sqrt{f^2(s)-f^2(s_0)}} ds & = & (1 + O(\tilde{s})) \frac{2b_j}{a_k} \int_{s_0}^{\tilde{s}} \frac{s^j}{s^k} \; \frac{s_0^k ds}{\sqrt{s^{2k}-s_0^{2k}}} \nonumber \\
     & = & (1 + O(\tilde{s})) \frac{2b_j}{a_k} 
s_0^{j+1-k} \cdot \int_1^{\tilde{s}/s_0} \frac{dy}{y^{2k-j} \sqrt{1 - y^{-2k}}} \nonumber \\
     & = & (1 + O(\tilde{s})) \frac{2b_j}{a_k} s_0^{j+1-k} \cdot C_{2k-j,2k}(\tilde{s}/s_0)
\end{eqnarray}
Substituting the behaviour of $C_{2k-j,2k}$ and taking $\tilde{s}$
fixed we get $l = O(s_0^k)$ for $k < (j+1)/2$, $l \sim -s_0^k \log s_0$
for $k = (j+1)/2$, and $l \sim s_0^{j+1-k}$ for $k > (j+1)/2$. To get
a more precise result in the latter case and prove assertion
\ref{l:eq0gt}, 
we can now take $\tilde{s} = s_0^{(2k-j-1)/(2k-j)}$.

\vspace{12pt}
Now we turn to investigate the behaviour of $K(s_0)$ as $s_0 \rightarrow 0$.
In order for it to be defined in the first place, we need two additional
conditions. One of them ensures the convergence of the second integral of 
(\ref{Ks0}), and the other ensures the convergence of the first integral
even for $s_1 = \infty$.

\begin{theorem}
\label{technical2}
Let $f(s),g(s)$ be functions as in theorem \ref{technical1}. Assume also that
\begin{enumerate}
\item \label{l2:zeroint} $j > -1$.
\item \label{l2:inftyint} $\int^\infty g(s)/f^2(s) ds < \infty$ 
(i.e the integral converges when its upper limit is taken to infinity).
\end{enumerate}
Take any $0 < s_1 \le \infty$, and define 
$K(s_0),\kappa$ as in (\ref{Ks0}).

\begin{enumerate}
\item \label{l2:f0ne0} if $f(0) \ne 0$, then we have  
$0< \kappa < \infty$, 
and as $s_0$ tends to $0$ from above, 
\begin{enumerate}
\item \label{K:ne0lt} if $k < 2(j+1)$, then $K(s_0) = \kappa + O(s_0^k)$.
\item \label{K:ne0eq} if $k = 2(j+1)$, then 
$K(s_0) = \kappa - b_j \sqrt{a_k/2f(0)} s_0^k \log s_0 
                                       + O(s_0^k \log (-\log s_0))$.
\item \label{K:ne0gt} if $k > 2(j+1)$, then 
$K(s_0) = \kappa + b_j \sqrt{a_k/2f(0)} ( C_{k/2-j,k}(\infty) + \frac{2}{k/2+j+1}) s_0^{k/2+j+1} + O(s_0^{k/2+j+1 + \frac{k/2-j-1}{k/2-j}})$.
\end{enumerate}
\item \label{l2:f0eq0} if $f(0)  =  0$, then we have that 
as $s_0$ tends to $0$ from above, 
\begin{enumerate}
\item \label{K:eq0lt} if $k < (j+1)/2$, then $K(s_0) = O(s_0^{2k})$.
\item \label{K:eq0eq} if $k = (j+1)/2$, then 
$K(s_0) = -{1 \over 2} b_j s_0^{2k} \log s_0 + O(s_0^{2k} \log (-\log s_0))$.
\item \label{K:eq0gt} if $k > (j+1)/2$, then 
$K(s_0) = b_j (\frac{1}{2} C'_{2k-j,2k}(\infty) + \frac{1}{j+1}) s_0^{j+1} 
                                       + O(s_0^{j+1 + \frac{2k-j-1}{2k-j}})$.
\end{enumerate}
\end{enumerate}
\end{theorem}
{\rm Proof:} We choose some $\tilde{s}$ such that the expansions 
(\ref{expansionf},\ref{expansiong}) are valid for $0 \le s \le \tilde{s}$.
We also note that for $0 \le x \le 1$, we have $(1 - \sqrt{1-x}) \sim x$.

The limit $s \rightarrow 0$ in the integrals for $K(s_0)$ is 
$\sim \int_0 g(s) ds < \infty$ by condition \ref{l2:zeroint}.
We shall now show that the first integral 
also converges in its upper limit even for $s_1 = \infty$.
The ratio of the integrand and the one of condition \ref{l2:inftyint} is 
\be
f^2(s) \left( 1 - \sqrt{1 - \frac{f^2(s_0)}{f^2(s)}} \right) \sim
f^2(s) \cdot \frac{f^2(s_0)}{f^2(s)} = f^2(s_0).
\ee
As the ratio is bounded, $K(s_0)$ (and $\kappa$) also converge.

Now let us look at
\begin{eqnarray}
\label{difK}
K(s_0) - \kappa  & =  & K(s_0) - K(0) \nonumber \\ 
        & = & \int_{s_0}^{s_1} \frac{g(s)}{f(s)} \left(\sqrt{f^2(s) - f^2( 0 )}- 
                                         \sqrt{f^2(s) - f^2(s_0)}  \right) ds \nonumber \\
                                  & &
 \mbox{} + \int_0^{s_0} \frac{g(s)}{f(s)} \sqrt{f^2(s) - f^2(0)} ds \\
        & \equiv & \Delta K_1(s_0) + \Delta K_2(s_0)
\end{eqnarray}
Although $\kappa \equiv K(0) = 0$ when $f(0)=0$, the above relation will
be useful in that case also. 

In order to evaluate $\Delta K_1(s_0)$, we shall divide its integration 
range into two parts. In the first one,
\begin{equation}
 \int_{\tilde{s}}^{s_1} \frac{g(s)}{f(s)} 
\left(\sqrt{f^2(s) - f^2( 0 )} - \sqrt{f^2(s) - f^2(s_0)}\right) ds  
\end{equation}
\vspace{-24pt}
\begin{eqnarray}
  & = &\int_{\tilde{s}}^{s_1} g(s) 
\left(\sqrt{1 - \frac{f^2( 0 )}{f^2(s)}} - 
      \sqrt{1 - \frac{f^2(s_0)}{f^2(s)}}\right) ds  \\
  & \sim & \int_{\tilde{s}}^{s_1} g(s) \frac{f^2(s_0) - f^2(0)}{2 f^2(s)} ds \\ 
  &  =   & \frac{f^2(s_0) - f^2(0)}{f^2(\tilde{s}) - f^2(0)} \cdot 
      \int_{\tilde{s}}^{s_1} g(s) \frac{f^2(\tilde{s}) - f^2(0)}{2 f^2(s)} ds \\
  & \sim & \frac{f^2(s_0) - f^2(0)}{f^2(\tilde{s}) - f^2(0)} \cdot  
          \int_{\tilde{s}}^{s_1} \frac{g(s)}{f(s)} 
\left(\sqrt{f^2(s) - f^2( 0 )} - \sqrt{f^2(s) - f^2(\tilde{s})}\right) ds \\ 
  & = & \frac{f^2(s_0) - f^2(0)}{f^2(\tilde{s}) - f^2(0)} \cdot 
        \Delta K_1(\tilde{s})
\end{eqnarray}
Therefore we see that the contribution of that 
part is $\sim (s_0/\tilde{s})^k    \Delta K_1(\tilde{s})$ if $f(0) \ne 0$, 
and     $\sim (s_0/\tilde{s})^{2k} \Delta K_1(\tilde{s})$ if $f(0)  =  0$.

In order to evaluate the other terms, 
we will have to deal separately with the cases $f(0) \ne 0$ and $f(0) = 0$. 
First we deal with the former case.  The lower part of $\Delta K_1(s_0)$ is
\[
\int_{s_0}^{\tilde{s}} \frac{g(s)}{f(s)} 
\left(\sqrt{f^2(s) - f^2( 0 )} - \sqrt{f^2(s) - f^2(s_0)}\right) ds 
\]
\vspace{-24pt}
\begin{eqnarray*}
& = & (1+O(\tilde{s})) \frac{b_j}{f(0)} \sqrt{2 f(0) a_k} 
\int_{s_0}^{\tilde{s}} s^j (\sqrt{s^k} - \sqrt{s^k - s_0^k}) ds \\
& = & (1+O(\tilde{s})) b_j \sqrt{2a_k/f(0)} s_0^{k/2+j+1} 
\int_1^{\tilde{s}/s_0} \frac{(1 - \sqrt{1 - y^{-k}})\;dy}{y^{-(k/2+j)}} \\
& = & (1+O(\tilde{s})) b_j \sqrt{2a_k/f(0)} s_0^{k/2+j+1} \cdot {1 \over 2} C'_{k/2-j,k}(\tilde{s}/s_0)
\end{eqnarray*}
Taking $\tilde{s}$ fixed and using lemma \ref{Cbehaviour}, we find
that $\Delta K_1(s_0) = O(s_0^k)$           for $k < 2(j+1)$, 
     $\Delta K_1(s_0) \sim -s_0^k \log s_0$ for $k = 2(j+1)$,
and  $\Delta K_1(s_0) \sim s_0^{k/2+j+1}$   for $k > 2(j+1)$.

$\Delta K_2(s_0)$ in the case $f(0) \ne 0$ is 
\begin{eqnarray*}
\int_0^{s_0} \frac{g(s)}{f(s)} \sqrt{f^2(s) - f^2(0)} ds & = &  
(1 + O(s_0)) \frac{b_j}{f(0)} \sqrt{2 f(0) a_k} \int_0^{s_0} s_0^{k/2+j} ds \\
& = & b_j \sqrt{2a_k/f(0)} \frac{1}{k/2+j+1} s_0^{k/2+j+1} + O(s_0^{k/2+j+2})
\end{eqnarray*}
Combining all the terms we demonstrate assertion \ref{K:ne0lt}. Letting 
$\tilde{s}$ vary with $s_0$, and using again lemma \ref{Cbehaviour} 
we can sharpen our results. By taking $\tilde{s} = -1/\log s_0$ we
demonstrate assertion \ref{K:ne0eq}, 
and by taking $\tilde{s} = s_0^{(k/2-j-1)/(k/2-j)}$ we demonstrate
assertion \ref{K:ne0gt}. 

Next we move to the case $f(0) = 0$.
The lower part of the integral of $\Delta K_1(s_0)$ is now
\begin{eqnarray}
\int_{s_0}^{\tilde{s}} \frac{g(s)}{f(s)} 
    \left(f(s) - \sqrt{f^2(s) - f^2(s_0)}\right) ds & = & \\
(1+O(\tilde{s})) b_j \int_{s_0}^{\tilde{s}} s^j 
    \left(1 - \sqrt{1 - (\frac{s_0}{s})^{2k}} \right) ds & = & \\
(1+O(\tilde{s})) b_j s_0^{j+1} \int_{1}^{\tilde{s}/s_0} y^j \left(1 - \sqrt{1 - y^{-2k}} \right) dy & = & \\
(1+O(\tilde{s})) b_j s_0^{j+1} \frac{1}{2} C'_{2k-j,2k}(\tilde{s}/s_0)
\end{eqnarray}
which gives,upon taking $\tilde{s}$ fixed, 
that $\Delta K_1(s_0) = O(s_0^{2k})$           for $k < (j+1)/2$, 
     $\Delta K_1(s_0) \sim -s_0^{2k} \log s_0$ for $k = (j+1)/2$,
and  $\Delta K_1(s_0) \sim s_0^{j+1}$          for $k > (j+1)/2$.

$\Delta K_2(s_0)$ is now very simple
\begin{eqnarray*}
\int_0^{s_0} \frac{g(s)}{f(s)} \sqrt{f^2(s) - f^2(0)} ds & = &  
\int_0^{s_0} g(s) ds \\
& = & (1 + O(s_0)) b_j \int_0^{s_0} s^j ds \\
& = & b_j {1 \over {j+1}} s_0^{j+1} + O(s_0^{j+2}) 
\end{eqnarray*}
Combining all the terms we demonstrate assertion \ref{K:eq0lt}.
In order to demonstrate assertion \ref{K:eq0eq} we use lemma \ref{Cbehaviour}
while taking $\tilde{s} = -1/\log s_0$, and for assertion \ref{K:eq0gt} 
we should be taking $\tilde{s} = s_0^{(2k-j-1)/(2k-j)}$.

\subsection{Asymptotics of $E(l)$}

In the computation of the behaviour of $E$ as a function of $l$, 
we shall take the limit $s_1 \rightarrow \infty$, so that strictly
speaking there are no "quarks" --- they are the limits of the
geodesic line. The condition ensuring that $K(s_0)$ does not diverge in this
limit will ensure also that $l$ does not diverge.

The geodesic line, coming from $s = \infty$, can not "pass the first 
valley of $f(s)$", or "climb a cliff of $g(s)$". 
We know, from theorem \ref{technical1}, what to demand in order that $l$ 
will diverge for $s_0 \rightarrow 0$. In order for $l$ not to diverge 
{\em before} that, we demand that $f(s)$ is increasing, (and therefore 
has no minimum for $s > 0$). 

After motivating our demands from $f(s),g(s)$, we are ready to
present the main result of the article:

\begin{theorem}
\label{main}
Let $\LL$ be as in (\ref{Lgeneral}), with functions $f(s),g(s)$ such
that:
\begin{enumerate}
\item \label{smoothf} $f(s)$ is analytic   for $0 < s < \infty$.
At $s = 0$, its expansion is:
\be
\label{expansion}
f(s) = f(0) + a_k s^k + O(s^{k+1})
\ee 
with $k > 0 \;,\; a_k > 0$.
\item \label{smoothg} $g(s)$ is smooth for $0 < s < \infty$. At $s = 0$,
its expansion is:
\be
g(s) = b_j s^j + O(s^{j+1})
\ee
with $j > -1 \;,\; b_j > 0$.
\item \label{positive} $f(s),g(s) \ge 0$ for $0 \le s < \infty$.
\item \label{increasing} $f'(s) > 0$ for $0 < s < \infty$.
\item \label{inftyint} $\int^\infty g(s)/f^2(s) ds < \infty$.

\end{enumerate}
Then for (large enough) $l$ there will be an even geodesic line asymptoting
from both sides to $s = \infty$, and $x = \pm l/2$.
As for the potential (\ref{Egeneral}) related to that configuration,
\begin{enumerate}
\item \label{conf}   if $f(0) > 0$, then
\begin{enumerate}
\item if $k = 2(j+1)$, $E = f(0) \cdot l -2 \kappa + 
                                    O((\log l)^\beta  e^{-\alpha l})$
\item if $k > 2(j+1)$, $E = f(0) \cdot l -2 \kappa - d \cdot l^{-\frac{k+2(j+1)}{k-2(j+1)}} + O(l^{-\frac{k+2(j+1)}{k-2(j+1)} - \frac{1}{k/2-j}}) $.
\end{enumerate}
with some $\beta$ and the positive constants
\begin{eqnarray*}
\kappa & = & \int_{0}^{\infty} \frac{g(s)}{f(s)}
\left(f(s) - \sqrt{f^2(s)-f^2(0)} \right) ds  \\
\alpha & = & \frac{\sqrt{2 f(0) a_k} \; k}{2 b_j} \\
d & = & \frac{2b_j}{k/2+j+1} \sqrt{\frac{a_k}{2f(0)}} \left(\frac{2b_j}{\sqrt{2f(0)a_k}} C_{k/2-j,k}(\infty) \right)^{\frac{k+2(j+1)}{k-2(j-1)}} 
\end{eqnarray*}
In particular, there is linear confinement.
\item \label{noconf} if $f(0) = 0$, then if $k > j+1$,
$E = -d' \cdot l^{-\frac{j+1}{k-j-1}} + O(l^{-\frac{j+1}{k-j-1} - \frac{2k-j-1}{(2k-j)(k-j-1)}})$ with
\be
d' = 2b_j \left(\frac{2b_j}{a_k} C_{2k-j,2k}(\infty)\right)^{\frac{j+1}{k-j-1}} D_{2k-j,2k}
\ee
In particular, there is no confinement.
\end{enumerate}
\end{theorem}
{\rm Proof:} Take any $s_0$ and choose arbitrary $\tilde{s} > s_0$.
The ratio between the integrand of $l$ (\ref{lgeneral}), 
and that of condition \ref{inftyint} is 
$1/\sqrt{1 - (f^2(s_0)/f^2(s))} $. As $f(s)$ is 
monotone increasing (condition \ref{increasing}), we have
\be
1 < 1/\sqrt{1 - (f^2(s_0)/f^2(s))} < 1/\sqrt{1 - (f^2(s_0)/f^2(\tilde{s}))}
\;\;\;\;\;\; \mbox{for} \;\; s > \tilde{s}
\ee
In other words, that ratio is $\sim 1$. 
As the integral in \ref{inftyint} converges,
so does that of (\ref{lgeneral}), and $l$ is finite when 
$s_1 \rightarrow \infty$. Hence, theorem \ref{technical1} is applicable. 

In all the cases we are going to prove, in particular $E < 0$ and so   
by theorem \ref{canoe}, $s_0$ decreases when $l$ increases,
and therefore converges at that limit.
We should show that the case $s_0 \rightarrow s^* \ne 0$ is impossible.
We can apply theorem \ref{technical1} with any $s = s^*$
instead of $s=0$. in the case $s^* \ne 0$, we have $k=1 , j \ge 0$ and 
therefore $k < 2(j+1)$, so by that theorem, $l \not \rightarrow \infty$.

Let us look now at assertion \ref{conf}. From (\ref{Egeneral}) we get
\be
E = f(0) \cdot l -2\kappa + \Delta E
\ee
with 
\be
\Delta E = (f(s_0)-f(0)) l - 2 (K(s_0) - \kappa)
\ee
We now evaluate 
$\Delta E$, first for the case $k = 2(j+1)$.   
From theorem \ref{technical1}, 
\be
l \le -\frac{2 b_j}{\sqrt{2 f(0) a_k}} \log s_0 + O(\log \log l)
\ee
 so $s_0 = O((\log l)^\gamma e^{-(\alpha/k) l})$ for some $\gamma$. 
From theorem \ref{technical2} we get
\be
K(s_0)-\kappa = \frac{a_k}{2} l s_0^k + O(\log l \;  s_0^k)
\ee
On the other hand, 
\be
(f(s_0)-f(0)) l = a_k l s_0^k  +O(l s_0^{k+1})
\ee
Therefore we have a cancelation, and 
\be
\Delta E = O(\log l \; s_0^k) = O((\log l)^\beta e^{-\alpha l})
\ee
as needed, with $\beta = \gamma k + 1$.

Now we turn to evaluate $\Delta E$ in the case $k > 2(j+1)$. Again we use
the functional dependence $l(s_0)$ from theorem \ref{technical1}, and get
\be
(f(s_0)-f(0)) l = 2 b_j \sqrt{\frac{a_k}{2f(0)}} C_{k/2-j,k}(\infty) s_0^{k/2+j+1} + O(s_0^{k/2+j+1 + \frac{k/2-j-1}{k/2-j}})
\ee
We also revert that functional dependence to get
\be
s_0 = \left(\frac{2b_j}{\sqrt{2f(0)a_k}} C_{k/2-j,k}(\infty) \right)^{\frac{1}{k/2-j-1}} l^{-\frac{1}{k/2-j-1}} +
O(l^{-\frac{1}{k/2-j-1} - \frac{1}{k/2-j}})
\ee
Using the value of $K(s_0)$ from theorem \ref{technical2}, we have a partial
cancelation, and we get 
\be
\Delta E = -\frac{2b_j}{k/2+j+1} \sqrt{\frac{a_k}{2f(0)}}
s_0^{k/2+j+1} + O(s_0^{k/2+j+1 + \frac{k/2-j-1}{k/2-j}})
\ee
Substituting $s_0$ we get the desired result.

Now we prove assertion \ref{noconf}. 
From theorem \ref{technical1}, we get 
\be
s_0 = \left(\frac{2b_j}{a_k} C_{2k-j,2k}(\infty)\right)^{\frac{1}{k-j-1}} 
l^{-\frac{1}{k-j-1}} + O(l^{-\frac{1}{k-j-1} - \frac{2k-j-1}{(2k-j)(k-j-1)}})
\ee
and 
\be
f(s_0) \cdot l = 2b_j C_{2k-j,2k}(\infty) s_0^{j+1} + O(s_0^{j+1 + \frac{2k-j-1}{2k-j}})
\ee
Using theorem \ref{technical2}, we get 
\be
E = 2b_j(C_{2k-j,2k}(\infty) - \frac{1}{2} C'_{2k-j,2k}(\infty) -
\frac{1}{j+1}) s_0^{j+1} + O(s_0^{j+1 + \frac{2k-j-1}{2k-j}}) 
\ee
which gives the desired result.

\section{Applications and verifications of the general result}

\subsection{invariance under reparameterizations of $s$}
The results obtained should be invariant under reparameterizations of $s$.
For example, if the "quarks" are supposed, in a certain setup, to reside
at $s_1 \ne \infty$, we can reparameterize $s$ so as to "move" them to 
$s_1 = \infty$. The simplest reparameterization is a dilatation, that is,
taking $s \mapsto \lambda s$, with $\lambda \ne 0$ a constant.
If, under this dilatation, a magnitude is multiplied by $\lambda^c$, 
we can say that this magnitude has conformal dimension $c$. It is easy to find 
the conformal dimensions of various magnitudes we have encountered. For example:
\begin{center}
\begin{tabular}{||l||c|c|c|c|c|c|c|c||}
\hline
magnitude & s   & x   & $\LL$ & $f(s)$ & $a_k$ & $g(s)$ & $b_j$  & $\alpha$  \\
\hline    
dimension & $1$ & $0$ & $0$   & $0$    & $-k$  & $-1$   & $-j-1$ & $j+1-k/2$ \\
\hline
\end{tabular}
\end{center}
We see that $\alpha$ has conformal dimension $0$, as it should, only when 
$k = 2(j+1)$. This is precisely when we have found the exponential correction 
to occur.

\subsection{General considerations}
When $f(0) > 0$, we see that a term proportional to $l^{-1}$ 
(of the order of the quantum correction L\"{u}scher term \cite{Luscher}) 
can not arise from that correction with $j > -1$, 
but that it is the limiting case as $k \rightarrow \infty$. The classical
correction computed in this article is always smaller, for large $l$, than
a $l^{-1}$ correction.

When $f(0) = 0$,  For the generic case $k = 2, j = 0$
we get a "Coulomb" potential $\sim l^{-1}$.
This case agrees with the potential found in \cite{Mal2} for 
${\cal N} = 4, D = 4$ SYM. Indeed, that theory is conformal and has no natural
length scale, so the potential must be $\sim l^{-1}$.

In the case where the correction to the potential is proven to be 
exponentially small, that is $O((\log l)^\beta e^{-\alpha l})$, we know of no 
explicit computation exhibiting that behaviour. 
In the explicit computations, $\lambda = O(1)$ always, and therefore 
$s_0^k \sim e^{-\alpha l}$. Moreover, the corrections to the behaviour
of $K(s_0)$ are then $O(s_0^k)$ and not $O(\log l \; s_0^k)$, and therefore 
$\Delta E$ and the corrections to the potential are $\sim e^{-\alpha l}$ 
or smaller. It may be that the $(\log l)^\beta$ is an artifact of the
proof. On the other hand, the cancelation of the 
$\sim l s_0^k = O((\log l)^\beta l e^{-\alpha l})$ term (which is always
$ O(l e^{-\alpha l})$ in the explicit calculations), shown in the proof of 
theorem \ref{main}, is generic, and was indeed observed in those
calculations, where it was considered accidental.
 
\subsection{The $AdS_5\times S^5$ dual of the 
${\cal N}=4$ SYM in four dimensions}
In this context, it is customary to use $U$ instead of $s$. The Lagrangian is
\cite{Mal2}
\begin{equation}
\label{LMaldacena}
\LL = \frac{1}{2\pi} \sqrt{U^4/R^4 + (U')^2}
\end{equation}
We see that $R$ is dimensionless and so the theory has no natural length scale.
From  (\ref{LMaldacena}) we extract $f(U) = (2\pi)^{-1} U^2/R^2$ and 
$g(U) = (2\pi)^{-1}$. Therefore,
$f(0) = 0 , k=2 , a_k = (2\pi)^{-1} R^{-2}$ and $j=0 , b_j = (2\pi)^{-1}$.
Moreover, $\int^\infty g(U)/f^2(U) dU \sim \int^\infty U^{-4} dU < \infty$.
Therefore, we can apply theorem \ref{main}. 
\be
C_{4,4}(\infty) = \frac{\sqrt{2}\pi^{3/2}}{\Gamma(\frac{1}{4})^2}
\ee
and so 
\be
d' = \frac{2\sqrt{2\pi}R^2}{\Gamma(\frac{1}{4})^2} D_{4,4}
\ee
and finally 
\be
E = - \frac{2\sqrt{2\pi}R^2}{\Gamma(\frac{1}{4})^2} D_{4,4} \cdot l^{-1} + 
O(l^{-7/4})
\ee
This result agrees completely with Maldacena's result \cite{Mal2}.

\subsection{Non-conformal cases with sixteen supersymmetries} 
A generalization of the former $D_3$ brane 
 Lagrangian to  $D_p$-branes  ($p \le 4$) with sixteen supersymmetries  
can be achieved following the steps taken in \cite{IMSY}.
For those cases the Lagrangian is given by 
\be
\LL = \frac{1}{2\pi} \sqrt{(U/R)^{7-p} + (U')^2}
\ee
With $R^{7-p}$ having the dimension of $\mbox{mass}^{3-p}$.
When $p \ne 3$, the theories are not conformal, so we do not expect a 
"Coulomb" potential. Indeed, $f(U) = (2\pi)^{-1} (U/R)^{(7-p)/2}$, 
so $f(0) = 0 , k=(7-p)/2 , a_k = (2\pi)^{-1} R^{-(7-p)/2}$, while 
$g(U) , j, b_j$ remain as in the former sub--section. 
Also, $\int^\infty g(U)/f^2(U) dU \sim \int^\infty U^{p-7} dU < \infty$. 
Hence,
$E = -d' \cdot l^{-2/(5-p)} +O(l^{-2/(5-p) - 2(6-p)/(5-p)(7-p)})$, with 
$d' \propto R^{(7-p)/(5-p)}$.
This result agrees with the computation for $p=2$, also performed in 
\cite{Mal2}.

\subsection{Dual models of pure YM theory in three and four dimensions}
Following a proposal of Witten \cite{Witads2} 
one can write down a gravity solution that corresponds to a pure YM theory.
For instance, to get  $YM_3$ one starts with the near extremal $D_3$ solution
in the near horizon limit and compactifies the  
Euclidean time direction on a circle. 
Upon taking the Wilson loop along this circle and along a space direction
one ends up with  a  scenario describing the four dimensional theory 
at finite temperature \cite{BISY2, RTY}. However, 
for Wilson loops along two space directions 
the limit of vanishing radius corresponds to a pure 
Euclidean YM theory in three dimensions.  
The string action takes, for that case, the following form
\be
\LL = \frac{1}{2\pi} \sqrt{(U/R)^4 + (U')^2 (1-(U_T/U)^4)^{-1}}
\ee
where the critical point (where $g(U)$ diverges) is $U=U_T$ and not $U=0$.
There,
$f(U_T) = \frac{1}{2\pi} (U_T/R)^2 \ne 0, k=1, a_k = \frac{U_T}{\pi R^2}$, 
while 
$g(U) = \frac{1}{2\pi} (1-(U_T/U)^4)^{-1/2}$ and
so $j=-1/2, b_j = \sqrt{U_T}/{4\pi}$. 
Also, $\int^\infty g(U)/f^2(U) dU \sim \int^\infty U^{-4} < \infty$.  
As $k = 2(j+1)$, we get from theorem \ref{main} that
$E = \frac{U_T^2}{2\pi R^2} \cdot l -2\kappa + O((\log l)^\beta e^{-\alpha l})$
with $\alpha = 2U_T/R^2$. 
A  detailed computation in \cite{GrOl} agrees with
the leading term, and does not include the constant one 
(due to a different subtraction scheme).
However, the next correction in \cite{GrOl} is claimed  to be  
$\sim l e^{-\alpha l}$. We believe that this is an erroneous result
\cite{Olesen} and in fact the correction behaves like $e^{-\alpha l}$. 

In the case of $QCD_4$, 
\be
\LL = \frac{1}{2\pi} \sqrt{(U/R)^3 + (U')^2 (1-(U_T/U)^3)^{-1}}
\ee
so $f(U_T) = \frac{1}{2\pi} (U_T/R)^{3/2} \ne 0, k=1, a_k = \frac{3\sqrt{U_T}}{4\pi R^{3/2}}$ and $j=-1/2, b_j = \sqrt{U_T}/{2\sqrt{3}\pi}$. Now,
$\int^\infty g(U)/f^2(U) dU \sim \int^\infty U^{-3} < \infty$. 
We get 
$E = \frac{U_T^{3/2}}{2\pi R^{3/2}} \cdot l -2\kappa + O((\log l)^\beta e^{-\alpha l})$
when now $\alpha = U_T^{1/2}/2R^{3/2}$.


\subsection{Dual models of pure YM theory at finite temperature}
When the time coordinate  is compactified,
and the Wilson loop is along this direction (on top of one space direction)
the  corresponding theory is a four dimensional 
  theory at finite temperature. It was shown in \cite{BISY1} that  
\be
\LL = \frac{1}{2\pi} \sqrt{(U/R)^4 (1-(U_T/U)^4) + (U')^2}
\ee

At the line $U = U_T$, $f(s)$ becomes negative. It does not have a minimum 
there ($k = 1$), and $g(s)$ does not diverge (it is constant, $j = 0$).
As $k < 2(j+1)$, $l$ is bounded as $U_0$ approaches $U_T$, and nothing seems
to prevent the string from entering the unphysical region $U < U_T$. 
However, it is argued in \cite{BISY1} that for values of $l$ above 
a critical one, in which the string reaches some $U_c > U_T$, 
the energy of the geodesic string is positive. In agreement with our
general considerations, the physical solution is found to be 
two "bare" quarks, and the potential is zero. 
Evidently, the reasons for the string not to enter the unphysical
region are outside the scope of our model.

\subsection{Pure YM from rotating branes}
Dimensionally reducing the theory of rotating M-branes \cite{CORT}
leads to the Lagrangian 
\be 
\LL = \sqrt{C} \sqrt{ \frac{U^6}{U_0^4} \Delta + (U')^2 \frac{U^2 \Delta}
               {1-a^4/U^4 - U_0^6/U^6}}
\ee


It is easy to see that there is a singular point $U_\infty$ for which 
$g^2(U)$ has a simple pole, while $f^2(U)$ is strictly positive for
$U_\infty \le U$. Hence, we have $k=1, j=-1/2$ and by theorem \ref{main} 
we find linear confinement with exponentially small correction.

\subsection{MQCD}
\label{MQCD}
The QCD string for the M theory version of ${\cal N}=1, D=4$ Super Yang--Mills
is characterized by \cite{KSS1}
\be
\LL = 2 \sqrt{2\zeta} \sqrt{\cosh(s/R)} \sqrt{1+s'^2} 
\ee
With $R$ (the radius of the 11-th dimension), and $\zeta$, related to 
$\Lambda_{QCD}$. 
We have $f(s) = g(s) = 2 \sqrt{2\zeta} \sqrt{\cosh(s/R)}$, and so
we find $f(0) = 2 \sqrt{2\zeta} , k = 2, a_k = \sqrt{2\zeta}/2R^2$ and
$j = 0 , b_j = f(0)$. 
In this case, 
$\int^\infty g(s)/f^2(s) ds \sim \int^\infty e^{-s/2R} ds < \infty$. 
Again $k = 2(j+1)$, and therefore
$E = 2 \sqrt{2\zeta} \cdot l -2\kappa + O((\log l)^\beta e^{-\alpha l})$ with 
$\alpha = 1/\sqrt{2}R$. The exact expression, computed in \cite{KSS1},
agrees with that result 
\footnote{Note that $\kappa$ is defined differently in the two contexts.}, 
and even gives a better estimate for the exponential term, as 
$O(l e^{-2\alpha l})$. 
A term $\sim (\log l)^\beta e^{-\alpha l}$ with $\beta \neq 0$ appears
nowhere in the expansion. The cancelation of the $\sim e^{-\alpha l}$
term, which does appear, is, in a sense, "accidental".
The cancelation of the $O(l e^{-\alpha l})$ 
(or $O(\epsilon \log \epsilon)$ in the notations of \cite{KSS1}), however,
which seemed also accidental,
is generic, as explained above.

When the supersymmetry is broken, we have \cite{KSS1}
\be
\LL = 2 \sqrt{2\zeta} \sqrt{\cosh(s/Rc)+ \mu} \sqrt{1+s'^2} 
\ee
with $c \approx 1$ and $\mu$ a soft supersymmetry breaking parameter.
Now $b_j = f(0) = 2 \sqrt{2\zeta} \sqrt{1+ \mu}$, and 
$a_k = \sqrt{2\zeta}/2\sqrt{1+\mu}R^2$, but $k,j$ do not change for $\mu > -1$.
The changes in the string tension and the constant term of the explicit 
computation of $E$ agree with the general result.
Now $\alpha = 1/\sqrt{2}\sqrt{1+\mu}R$, and the "accidental"
cancelation of the $e^{-\alpha l}$ term persists, so the
exponential term in the explicit computation is shown to be again 
$O(l e^{-2\alpha l})$. Furthermore, that computation shows that 
the exponential correction can be either positive (for 
$\mu_1 < \mu < \mu_2$, with $\mu_{1,2} = 27 \mp 16 \sqrt{3}$) or negative.
For $\mu = \mu_{1,2}$ there is a further cancelation, and the correction
is $O(e^{-2\alpha l})$.
 
When $\mu = -1$, we have $f(0)=0 , k=j=1$. As $k = (j+1)/2$, $l$ does not
grow to infinity as $s_0$ approaches $0$, and theorem \ref{main} is not
applicable. As mentioned in \cite{KSS1}, above a certain value of $l$, 
the configuration of two bare quarks is energetically favourable to the
string obeying the Euler--Lagrange equations. (Very loosely speaking,
$D_{1,2} = -\infty$).
 
\subsection{Polyakov's non--critical string}
Polyakov suggests \cite{Polyakov} the conformal invariant solution
$a(\phi) = e^{\alpha \phi}$ with $\alpha$ constant. The range of
$\phi$ is $-\infty < \phi < \infty$ and the "quarks" are situated at
$\phi = \infty$. The function $f(\phi) = a^2(\phi)$ has no minimum at
a finite value of $\phi$, and $g(\phi) = a(\phi)$ does not diverge, so
our theorem \ref{main} is not directly applicable. We can, however,
reparameterize $\phi$ and set $s = a(\phi) = e^{\alpha \phi}$. Now,
\be
\LL = \sqrt{f^2(\phi) + g^2(\phi) \phi'^2} = 
      \sqrt{e^{4\alpha \phi} + e^{2\alpha \phi} \phi'^2} = 
      \sqrt{s^4 + \alpha^{-2} s'^2}
\ee 
and we are essentially back in Maldacena's case which is conformal, as we have
indeed seen.

\section{The even vs. odd cases}

If the functions $f(s),g(s)$ are defined also for $s < 0$, and are even
functions of $s$, then it makes sense to look also at the case where the
quark and anti--quark are situated at $s = \pm s_1$ and the string describes
an odd function $s(x)$, and to compare it to the previous case in which 
$s(x)$ is even.

The Lagrangian and Hamiltonian are equal to those of the previous
case, but, as $s(0) = 0$, the solution is now specified by the value of 
$r_0 \equiv s'(0)$, and
\begin{equation}
\HH(s,s') = \HH(0,r_0) = - \frac{f^2(0)}{\sqrt{f^2(0) + g^2(0) r_0^2}}
\end{equation}
(from which we see that we must have either $f(0) \ne 0$ or $ g(0) \ne 0$).
The equations for $\frac{ds}{dx} \;,\; l \;,\; E$ can be extracted,
and a treatment of this case, following the lines of the previous one, 
can be achieved. We shall not pursue this course. Instead, we shall give
a simple relation between the energies in the two cases.

\begin{figure}[h!]
\begin{center}
\resizebox{0.8\textwidth}{!}{\includegraphics{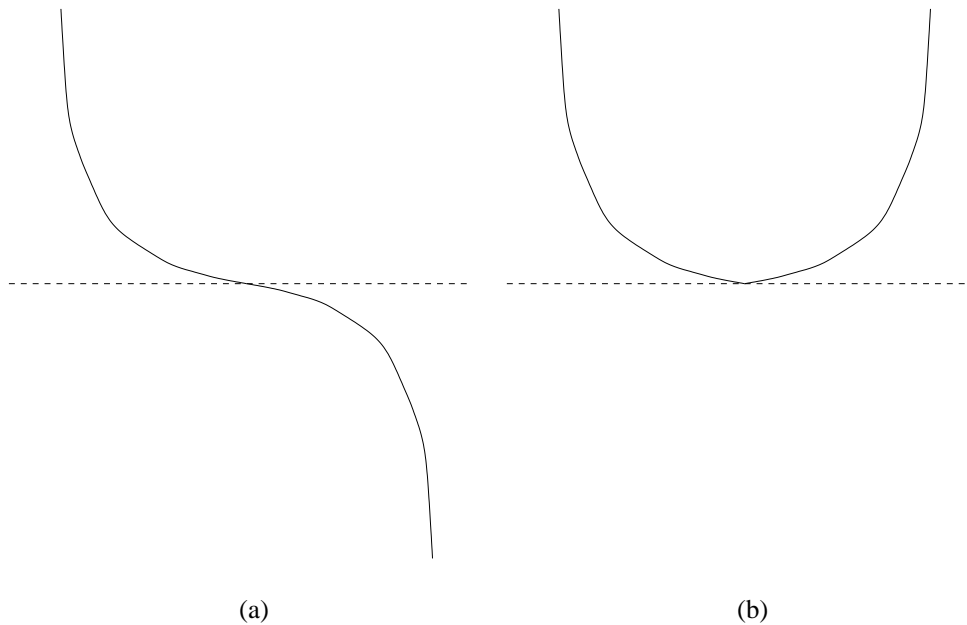}}
\end{center}
\caption{(a) The minimal energy odd string. 
         (b) The derived even string       }
\label{fig:asymsol}
\end{figure}

\begin{figure}[h!]
\begin{center}
\resizebox{0.8\textwidth}{!}{\includegraphics{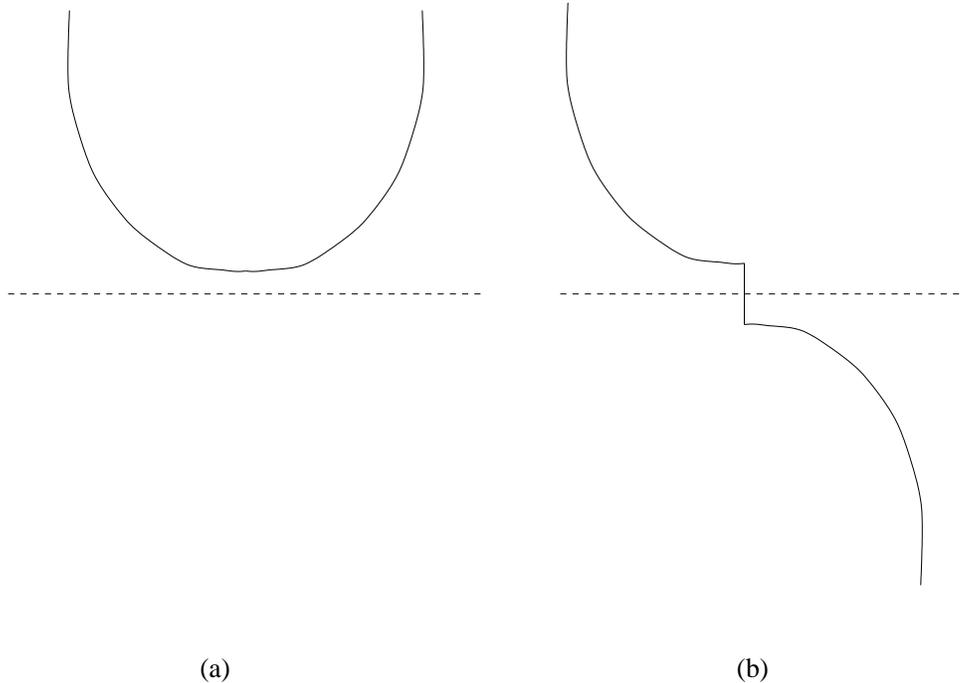}}
\end{center}
\caption{(a) The minimal energy even string. 
         (b) The derived odd string}
\label{fig:symsol}
\end{figure}

\begin{theorem}
\label{evenodd}
Let $E_{\mbox{even}}(l)$ be the quark anti--quark potential in the even case 
as a function of their separation, and let $E_{\mbox{odd}}(l)$ be the 
corresponding potential in the odd case. Let $s_0$ be the smallest value
of $s$ attained by the even string. Then,
\begin{equation}
0 \le E_{\mbox{odd}}(l) - E_{\mbox{even}}(l) \le 2 \int_0^{s_0} g(s) ds
\end{equation}
\end{theorem}
{\rm proof:} If we reflect, for negative $x$, the graph of the odd $s(x)$ 
string, we get an even one with the same energy (which is not in equilibrium 
for $x = 0 , s(x) = 0$). The solution for the even case has,
of course, a smaller energy. Therefore 
$E_{\mbox{odd}}(l) \ge E_{\mbox{even}}(l)$ (see figure (\ref{fig:asymsol})).
On the other hand, if we break the graph of the even $s(x)$ string 
into two parts, put them at different sides of $s=0$, and connect them 
with a straight string segment at $x=0, -s_0 \le s \le s_0$, as shown in
figure (\ref{fig:symsol}), we get a string describing an odd function 
(which is not in equilibrium for $x = 0 , s(x) = \pm s_0$).  
The solution for the odd case has a smaller energy, and hence
$E_{\mbox{odd}}(l) \le E_{\mbox{even}}(l) + \int_{-s_0}^{s_0} g(s) ds$.

This simple theorem is sufficient to give strong estimates on the difference
of the potentials in the two cases. Let us examine the MQCD setup discussed
in sub--section \ref{MQCD}. In that setup, 
$2 \int_0^{s_0} g(s) ds \approx 2\sqrt{2\zeta} \cdot 2 s_0 \sim R \sqrt{\zeta} e^{-(\alpha/2)l}$,
while the explicit computations \cite{KSS1} give 
$E_{\mbox{odd}}(l) - E_{\mbox{even}}(l) \sim R \sqrt{\zeta} e^{-\alpha l}$.
We see that theorem \ref{evenodd} gives an exponentially small estimation
of the difference, which is not too bad. It is easy to see that if the
correction is exponentially small for the even case, it always remains so 
also in the odd case.

\section{Summary and conclusions}

Recently, Wilson loops, or quark anti--quark potentials, were computed 
from various string actions \cite{Mal2,HSZ,KSS1,BISY1,RTY,BISY2}.
In this note we suggest a "unified" framework for discussing all those cases 
and others. The basic setup includes  
 a "quark" and an "anti--quark"  situated 
in flat space, connected by a
 string stretched in that space and in an extra 
curved dimension. The space--time metric depends only on this extra  
coordinate.
We have identified the cases in which the position of the middle of the string 
approaches a constant value of the extra dimension as the quark anti--quark 
separation $l$ grows, and have computed the potential in that limit from the 
leading terms of the Taylor expansion of the metric around that constant 
value. 
We have shown that the linear coefficient of the potential is equal to 
the string tension of a string situated
at the aforementioned constant value of the extra dimension. 
Therefore, confinement arises exactly when that string tension 
is non--vanishing.  In more technical terms,  we have shown that 
(assuming without loss of generality that $f(s)$ has a minimum or 
$g(s)$ diverges at $s=0$)
{ \it confinement occurs if and only if $f(0)>0$ and the 
corresponding string tension is  $f(0)$}. 

In case of confinement, we have shown that the correction to the linear
potential (apart from a constant term) is either a negative power 
of the separation, or exponentially small. In both cases we explicitly 
find the relevant constants.
The exponentially small correction
arises in the critical case when the 
minimum of $f(s)$ is just deep enough (or the divergence of $g(s)$ is just
strong enough) to allow the separation to diverge as the string approaches
the minimum. This case arises when the minimum of $f(s)$ at $s=0$ is quadratic,
and $g(s)$ neither diverges nor has a zero there, and therefore is the 
generic one.   

We have proven that the exponentially small correction
is $O((\log l)^\beta e^{-\alpha l})$, when we have identified $\alpha$
but not $\beta$. In all explicit calculations $\beta = 0$, and the
$(\log l)^\beta$ factor might be an artifact of the proof.
We detect a cancelation of the $\sim l  s_0^k$
terms (which are $\sim l e^{-\alpha l}$ in the explicit calculations).
This cancelation was
encountered in explicit computations  and we argue that 
it is in fact generic. 
This result contradicts the results  of \cite{GrOl} where it was claimed 
that the leading correction  is of the form  $l  e^{-\alpha l}$. 
Some of the explicit computations show that the true behaviour 
of the correction is $e^{-\alpha l}$, so our bound is rather tight.

When there is no confinement, the potential we find is asymptotically 
a negative power of the separation, and we find the exact power 
and coefficient.
In  particular, for the large $N$
CFT of \cite{Mal2},  we explicitly re-derive the potential, 
including the numerical constants.

We demonstrate our general  results by applying them to a  set of 
string configurations that 
were studied recently \cite{Mal2,HSZ,KSS1,BISY1,RTY,BISY2}. 

The next step in this program of "precision measurements" of the Wilson loops 
or the quark anti--quark potential is not to compute  higher order
corrections of the classical computation but rather to determine the 
quantum fluctuations. For a string that is  
stretched only 
along the flat four dimensions with no extra curved dimension,  
 the quantum fluctuations were computed 
in \cite{Luscher}. The result is a correction of the form ${c\over L}$ where
$c$ is a universal coefficient independent of the coupling constant. 
We are studying \cite{KSS3} a generalization of this construction to the 
string configurations discussed in this note.

\end{document}